\documentclass[sigconf,nonacm]{acmart}
\usepackage{epsfig}
\usepackage{graphicx}
\usepackage{dcolumn}
\usepackage{bm}
\usepackage{tikz}
\usepackage[all]{xy}
\usetikzlibrary{arrows,shapes}
\usepackage{centernot}

\usepackage{color} 

\AtBeginDocument{%
	\providecommand\BibTeX{{%
			\normalfont B\kern-0.5em{\scshape i\kern-0.25em b}\kern-0.8em\TeX}}}

\setcopyright{acmcopyright}
\copyrightyear{2020}
\acmYear{2020}
\acmDOI{10.1145/1122445.1122456}

\acmConference[ESEC/FSE 2020]{The 28th ACM Joint European Software Engineering Conference and Symposium on the Foundations of Software Engineering}{8 - 13 November, 2020}{Sacramento, California, United States}

\begin{document}
	
	\title{Topological Differential Testing}
	
	\author{Kristopher Ambrose}
	\affiliation{%
		\institution{BAE Systems FAST Labs}
		\streetaddress{4301 North Fairfax Drive}
		\city{Arlington}
		\state{Virginia}
		\postcode{22203}
	}
	\email{kristopher.ambrose@baesystems.com}
	
	\author{Steve Huntsman}
	\orcid{0000-0002-9168-2216}
	\affiliation{%
		\institution{BAE Systems FAST Labs}
		\streetaddress{4301 North Fairfax Drive}
		\city{Arlington}
		\state{Virginia}
		\postcode{22203}
	}
	\email{steve.huntsman@baesystems.com}
	
	\author{Michael Robinson}
	\affiliation{%
		\institution{American University}
		\city{Washington, DC}
	}
	\email{michaelr@american.edu}
	
	\author{Matvey Yutin}
	\affiliation{%
		\institution{BAE Systems FAST Labs}
		\streetaddress{4301 North Fairfax Drive}
		\city{Arlington}
		\state{Virginia}
		\postcode{22203}
	}
	\email{matvey.yutin@baesystems.com}
	
	\renewcommand{\shortauthors}{Ambrose, Huntsman, Robinson, and Yutin}
	
	\begin{abstract}
		We introduce \emph{topological differential testing} (TDT), an approach to extracting the consensus behavior of a set of programs on a corpus of inputs. TDT uses the topological notion of a simplicial complex (and implicitly draws on richer topological notions such as sheaves and persistence) to determine inputs that cause inconsistent behavior and in turn reveal \emph{de facto} input specifications. We gently introduce TDT with a toy example before detailing its application to understanding the PDF file format from the behavior of various parsers. Finally, we discuss theoretical details and other possible applications.
	\end{abstract}

	
	
	
	\maketitle
	

	\section{Introduction \label{sec:Intro}}
	
	\emph{Differential testing} (DT) \cite{mckeeman1998differential,gulzar2019perception} is an approach to software testing that evaluates the behavior of multiple versions of a program on multiple inputs. Frequently, DT is performed in the context of only two programs, one of which serves as an oracle. However, most programs do not have a formal specification or an oracle that evaluates the correctness of their behavior, and many of the most ubiquitous programs (e.g., browsers, document editors, viewers, etc.) have mutiple versions and/or analogues with broadly equivalent functionality. Indeed, \emph{$N$-version programming} \cite{avizienis1985n} elevates this to a software design principle. Meanwhile, formal verification of software frequently does not extend to tasks such as parsing inputs or unparsing outputs, or may rely upon unverified libraries which have multiple versions. For example, \texttt{Csmith} \cite{yang2011finding} found bugs in unverified parts of the mostly formally verified \texttt{CompCert 1.6} compiler.
	\footnote{
		The formally verified part of \texttt{CompCert 3.6} (the most recent version as of this writing) only extends from source abstract syntax tree to assembly abstract syntax tree--with one exception: the parser is also formally verified now \cite{jourdan2012validating}. However, a transformation stage from C to \texttt{CompCert} abstract syntax is still not covered by a formal proof \cite{leroy2016compcert}. On a related note, the problem of formally verifying a C11 parser is ill-posed due to ambiguities in the informal prose standard \cite{jourdan2017simple}.
	} 
	Finally, plug-ins and configurations (both internal and external) can introduce further variability into program behaviors.
	
	In such a context, we can extract a \emph{de facto} specification of acceptable inputs from the collective behavior of similar programs on a suitably representative corpus. A naive approach in this vein is to perform a simple voting procedure, which can also yield a \emph{de facto} specification of behavior if there is a clear consensus. This tactic works well for, e.g., independently produced compilers, since they do not exhibit correlated failures, presumably due to diversity among internal representations on which most of the compilation effort is concentrated \cite{yang2011finding}. However, compilers are atypical pieces of software: they have carefully designed parsers and intricate algorithm implementations. More typically, programs have \emph{ad hoc} parsers and comparatively simple (if also extensive) internal logic \cite{jana2012abusing,momot2016seven}. This explains the observation of \cite{knight1986experimental} that (non-compiler) software failures are correlated across versions, and it highlights the need for a collective differential testing approach to yield \emph{de facto} specifications.
	
	Towards this end, we introduce \emph{topological differential testing} (TDT), which exploits structure in topological representations of the collective behavior of a set of programs on a corpus of inputs. Specifically, TDT forms weighted abstract simplicial complexes and sheaves \cite{ghrist2014elementary} that encode inconsistencies and subsequently identifies the inputs (or via duality and as scale permits, programs) that behave inconsistently. TDT thereby distills an input corpus to a consistent subcorpus that reflects a \emph{de facto} input specification. 
	
	While DT has been used for compilers \cite{yang2011finding,barany2018finding}, parsers in cryptographic protocols such as X.509/SSL/TLS \cite{kaminsky2010pki,brubaker2014using,argyros2016sfadiff,tian2019differential}, browsers \cite{choudhary2013x,watanabe2019layout}, and (applications targeting) Android and Java virtual machines \cite{kyle2015application,chen2016coverage,fazzini2017automated,chen2019deep}, our own considerations are focused on parsers for complex formats or protocols such as PDF \cite{ISO32000,whitington2011pdf} or DDS/RTPS \cite{DDS,RTPS}. In practice, we run a corpus such as the PDF portion of \cite{garfinkel2010digital} through sandboxed parsers \cite{greamo2011sandboxing,maass2016systematic} to produce a relation between files and parsers that we then analyze using TDT. 
	
	This application is motivated by the fact that complex formats such as PDF have \emph{de facto} specifications in lieu of, or that substantially differ from, an informal (i.e., not machine-readable) standard or specification. Such cases are signified by \emph{parser differentials} \cite{kaminsky2010pki,jana2012abusing,momot2016seven}, where various parsers disagree on the validity of certain inputs.
	\footnote{
		NB. \cite{kuchta2018correctness} essentially applied DT to PDF readers and files at the rendering level by performing image processing and error message clustering. We actually employ a similar tactic in our work by using \texttt{Cuckoo Sandbox} to obtain rendered images and error messages, while also facilitating additional capabilities such as malware detection \cite{xu2016automatically}.
	}
	While DT frequently relies on domain-specific knowledge of inputs (see, e.g., \cite{brubaker2014using}), TDT assumes the existence of an appropriate reference corpus such as \cite{garfinkel2010digital} and a set of programs that together implicitly define an input specification. 
	\footnote{
		NB. A reference corpus can be augmented via differential fuzzing \cite{jana2012abusing, argyros2016sfadiff}, which essentially seeks to evolve an initial corpus into one whose distribution of accept/reject statistics tends to the uniform distribution on the power set of programs under consideration \cite{petsios2017nezha}. (For general background on fuzzing, see \cite{manes2019art}; for a state-of-the-art technique, see, e.g., \cite{she2019neuzz}.) However, this tactic seems to be fundamentally in tension with TDT.
	}
	
	In this paper we restrict consideration to programs that produce binary outputs: e.g., parsers, binary classifiers, etc. The binary output may simply be a side effect, e.g., the presence or absence of an error, or the result from a binary question about a more generic output.
	While part of the motivation for our more abstract mathematical considerations \S \ref{sec:Theory} is to pave the way for directly analyzing richer non-binary outputs, various postprocessing techniques can transform generic outputs into binary ones. For example, given a suitable (dis)similarity measure on outputs as in \cite{kuchta2018correctness}, we can cluster the outputs and label members of the largest cluster (with some tiebreaking mechanism) with a 1, and all other outputs with a 0, in order to obtain a binary relation suitable for TDT. This can be particularly useful in situations where unspecified or explicitly undefined behavior makes it possible for programs to produce different outputs from the same input without signaling a fault.
	
	The paper is structured as follows: \S \ref{sec:Motivation} provides motivation and a gentle introduction to the theoretical framework for TDT. \S \ref{sec:Examples} demonstrates TDT by way of a series of examples, and \S \ref{sec:Theory} provides theoretical details. Finally, we make general remarks in \S \ref{sec:Remarks}.

	\section{Motivation \label{sec:Motivation}}
	
\begin{figure}[htpb]
		\centering
		\includegraphics[trim = 50mm 110mm 45mm 105mm, clip, width=\columnwidth,keepaspectratio]{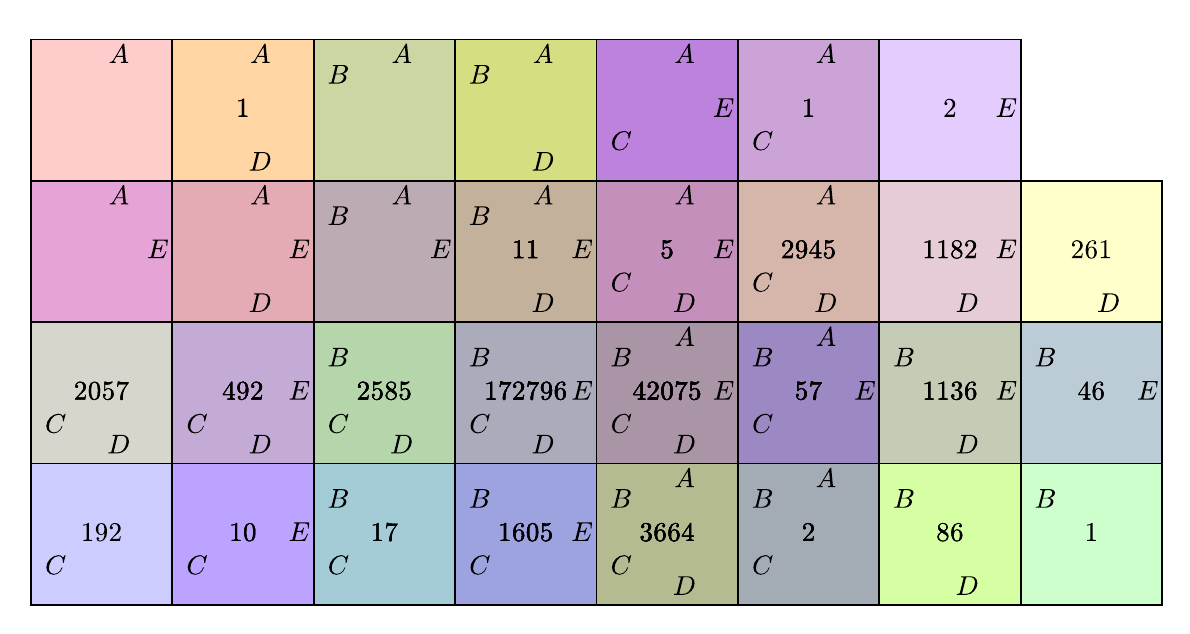}
		\caption{ \label{fig:polyvenn5} Visual interpretation of a Venn diagram becomes exceedingly difficult. Here, we show a Venn diagram for five sets $A, \dots, E$ corresponding to parsers, and in which all intersections (i.e., subsets of parsers) are unit squares (in the terminology of \cite{bultena2012minimum}, a ``$(2,3)$ polyVenn''). The diagram is labeled with weights corresponding to the numbers of files accepted by a given subset of parsers in the example of \S \ref{sec:Govdocs1}.
		}
	\end{figure}

\begin{figure}[htpb]
		\centering
		\includegraphics[trim = 50mm 110mm 45mm 105mm, clip, width=\columnwidth,keepaspectratio]{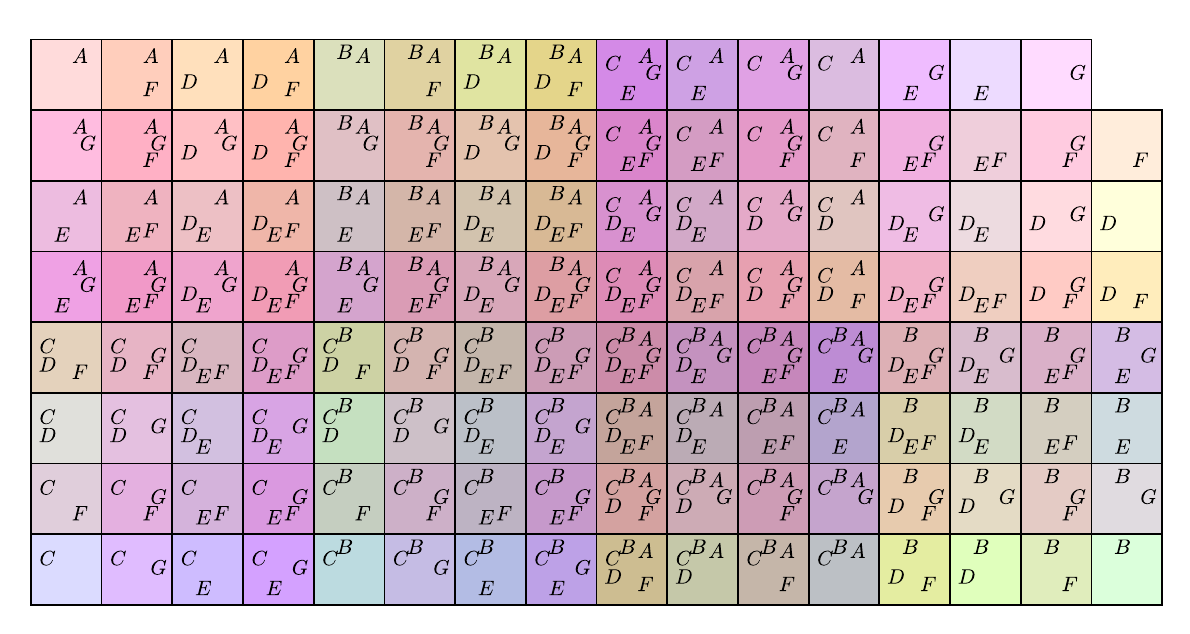}
		\caption{ \label{fig:polyvenn7} A $(3,4)$ polyVenn. Though state of the art at this scale, such a representation is functionally almost useless.
		}
	\end{figure}

	A \emph{de facto} file format is determined by the collective behavior of several parsers. A given file will be \emph{accepted} (successfully parsed) by some parsers, but \emph{rejected} by others.  By comparing the set of files that two parsers accept, we can tell whether one parser is more stringent than the other, or if the two parsers disagree entirely about which files should be accepted.  We advocate that one should consider the set of files accepted by a given \emph{set} of parsers.  For small sets of parsers or files, this kind of analysis can be done in an \emph{ad hoc} manner, though the bookkeeping can become quite complex for even five parsers, as Figure \ref{fig:polyvenn5} shows. While it might be argued that the particular Venn diagram in Figure \ref{fig:polyvenn5} is not optimal for human understanding, ``nice'' diagrams in which sets are represented as congruent ellipses only exist up to the case of five sets. For larger collections of sets (e.g., seven sets as in Figure \ref{fig:polyvenn7}), this sort of Venn diagram represents the state of the art, which evidently lags behind what we might desire.
	This complexity is effectively managed by the \emph{weighted Dowker complex}, whose formal definition we postpone until \S \ref{sec:Theory}.  We use the weighted Dowker complex to determine which files are potentially \emph{inconsistent} with a consensus \emph{de facto} format, and demonstrate that this notion of inconsistency can help cluster contentious aspects of the format.

	\subsection{Theoretical framework \label{sec:Framework}}
	
	Suppose that a given file $f$ is accepted by three parsers, $A$, $B$, and $C$. If we remove parser $C$ from consideration, this does not change the fact that parsers $A$ and $B$ still accept $f$.   The (unweighted) \emph{Dowker complex} records all sets of parsers that accept some file in common \footnote{Or some minimal number of files in common, should one obtain too many sets.}.  While it would make linguistic sense to say that $f$ is accepted by both the set $\{A,B,C\}$ and the set $\{A,B\}$ of parsers, this yields somewhat weak information.  What is important about differential testing is the \emph{differential}, after all!  To emphasize this point, let us only say that a given file $f$ is \emph{accepted by a subset of parsers} if this subset is maximal, relative to some ambient set of parsers.  Concretely, suppose that we are only concerned with four parsers $A$, $B$, $C$, and $D$.  If $D$ rejects $f$, then we shall say that $f$ is accepted by the set $\{A,B,C\}$, but not the set $\{A,B,C,D\}$ (since $D$ rejects $f$), nor the set $\{A,B\}$ (since this is not maximal).  The number of files (or even the set of files themselves) serve as the \emph{weights} on each subset of parsers.
	
	Let us now consider the difference between the sets of files accepted by two sets of parsers $X$ and $Y$.  Based on our definition, if we assume that $X \subseteq Y$ we cannot assume that the set of files accepted by $X$ is a subset of (or a superset of) the set of files accepted by $Y$.  But under the hope that there \emph{actually is} a \emph{de facto} format for the files being accepted or rejected by the parsers, we generally expect the set of files accepted by $Y$ will be larger than that of $X$.  That is to say, the number of files accepted by each parser in the set difference $Y \backslash X$ should be small if there is to be consensus between the parsers about the format.  Given this hope, we say that $X$ and $Y$ are \emph{consistent} if the number of files accepted by $Y$ is larger than that of $X$ and \emph{inconsistent} otherwise.  Intuitively, two sets of parsers are inconsistent if they disagree about the \emph{de facto} format.
	
	We can compute \emph{which} files are at fault for inconsistency between two sets $X \subseteq Y$ of parsers, simply by identifying which files accepted by every parser in $X$ are rejected by at least one parser in $Y$.  We call those files \emph{inconsistent for the pair of sets of parsers $X \subseteq Y$}.  Briefly, a file $f$ is inconsistent if $X$ is not the largest set of parsers in which each one accepts $f$.   A given file being inconsistent means that it potentially causes the parsers to disagree about the \emph{de facto} format, and the file therefore likely contains a contentious feature.
	
	Since there may be several points of contention between parsers about the \emph{de facto} format, it is wise to prioritize them.  Our strategy is to sweep over all pairs of subsets of parsers, recording how many deem a given file inconsistent, which we call the file's \emph{inconsistency score}.  The example in \S \ref{sec:Govdocs1} highlights that files with similar usage patterns of contentious features are clustered by the inconsistency score.  Entirely compliant files will typically have a low score, as will files with severe format problems.  On the other hand, files that have an unusually high inconsistency score will tend to show more subtle and varied issues.  This seems to happen because developers of relatively permissive parsers make minor -- but different -- assumptions about how to repair partially compliant files.

	\section{Examples \label{sec:Examples}}
	
	This section outlines three different example applications of TDT to identify files that fail to meet an \emph{ad hoc} consensus file format.
	\begin{enumerate}
		\item A toy example to outline the construction;
		\item The \texttt{Govdocs1} corpus \cite{garfinkel2010digital}, in which the format is mostly followed, and as processed by a set of five parsers;
		\item A test corpus produced under the aegis of the DARPA SafeDocs program, in which many files that violate the format are included.
	\end{enumerate}

	\subsection{A toy example \label{sec:toy}}
	
	Suppose that we have four parsers ($A, \dots, D$) that attempt to read $20$ files.  Each parser may either \emph{accept} or \emph{reject} each file, which we can encode as a binary $1$ or $0$, respectively.  It is helpful to organize all of the results in matrix form, such as
	\[R = \left(\begin{smallmatrix}
	1&0&0&0&0&0&1&1&1&1&1&1&0&0&0&0&1&1&1&1 \\
	0&1&1&0&0&0&1&1&1&0&0&0&1&0&0&0&0&0&0&0 \\
	0&0&0&1&1&0&0&0&0&1&0&0&1&1&1&1&1&1&1&1 \\
	0&0&0&0&0&1&0&0&0&0&1&1&0&1&1&1&1&1&1&1
	\end{smallmatrix}\right).\]
	in which each column represents a file and each row represents a parser.  For instance, the first file (column) was only accepted by the first parser; it was rejected by all other parsers.  Since there are four parsers, there are $16$ possible subsets of parsers including the empty set.  Let us collect which subsets of parsers accept at least one file in common.  For instance, parsers (rows) $A$, $C$, and $D$ all accept the last four files (columns).  Parsers $B$ and $C$ both accept file (column) $13$ in common; even though both parsers accept other files, they do not agree on accepting any other file.  There are also subsets of parsers that do not agree on accepting \emph{any} file.  For instance, parsers $A$, $B$, and $C$ cannot all agree on accepting any file, even though every pair of parsers out of these three accepts a file in common.  
	
	\begin{figure}[htbp]
		\centering
		\begin{tikzpicture}[->,>=stealth',shorten >=1pt,scale = 0.9,every node/.style={transform shape}]]
		\node [draw, align=center] (vA) at (-3,3) {$\{A\}$; $1$};
		\node [draw, align=center] (vB) at (-1,3) {$\{B\}$; $2$};
		\node [draw, align=center] (vC) at (1,3) {$\{C\}$; $2$};
		\node [draw, align=center] (vD) at (3,3) {$\{C\}$; $1$};
		\node [draw, align=center] (vAB) at (-4,1.5) {$\{A,B\}$; $3$};
		\node [draw, align=center] (vAC) at (-2,1.5) {$\{A,C\}$; $1$};
		\node [draw, align=center] (vAD) at (0,1.5) {$\{A,D\}$; $2$};
		\node [draw, align=center] (vBC) at (2,1.5) {$\{B,C\}$; $1$};
		\node [draw, align=center] (vCD) at (4,1.5) {$\{C,D\}$; $3$};
		\node [draw, align=center] (vACD) at (0,0) {$\{A,C,D\}$; $4$};
		\draw (vACD) to (vAC);
		\draw (vACD) to (vAD);
		\draw (vACD) to (vCD);
		\draw (vAB) to (vA);
		\draw (vAB) to (vB);
		\draw (vAC) to (vA);
		\draw [color=red] (vAC) to (vC);
		\draw (vAD) to (vA);
		\draw (vAD) to (vD);
		\draw [color=red] (vBC) to (vB);
		\draw [color=red] (vBC) to (vC);
		\draw (vCD) to (vC);
		\draw (vCD) to (vD);
		\end{tikzpicture}
		\caption{ \label{fig:toyComplex} Weighted Dowker graph for the toy data using ``reverse'' order ($\{\bullet\}$). Faces of the Dowker complex (see \S \ref{sec:Theory}) correspond to graph nodes, which are marked with tuples of rows followed by face counts. {\color{red}Inconsistent edges are red.}
		}
	\end{figure}
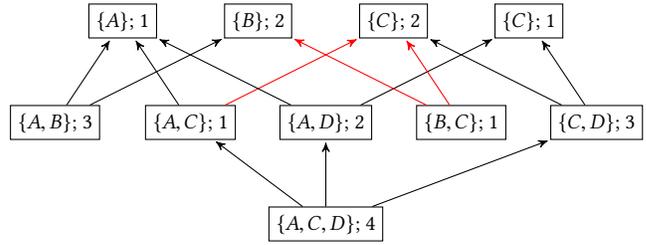
	
	It is helpful to organize the information collected in the matrix in a diagram called a \emph{weighted Dowker graph}, shown in Figure \ref{fig:toyComplex}.  The figure shows each subset of parsers that agrees on accepting at least one file.  The labels also include the number of common accepted files, so the label $\{A,C,D\}; 4$ means that there are $4$ files that are jointly accepted by each of the three parsers $A$, $C$, and $D$. We will call the number of files accepted by a subset of parsers the \emph{weight} of that subset. The arrows indicate subset relations, so that each subset of parsers points to each of \emph{its} subsets.  
	
	Recalling that we are looking for a consensus about an \emph{ad hoc} format, we expect that disagreements between parsers should be uncommon.  When we exclude a parser from consideration, any resulting subset of parsers should accept fewer files; weights should not increase when following the arrows in the weighted Dowker graph.  Every instance where this assumption holds is termed a \emph{consistent subset relationship} and is marked with a black arrow in Figure \ref{fig:toyComplex}; otherwise it is termed \emph{inconsistent} and is marked with a {\color{red}red} arrow.
	
	As noted before, parsers $A$, $C$ and $D$ all accept four files in common.  If we exclude parser $D$, then the resulting subset (parsers $A$ and $C$) additionally accepts file $10$ in common.  (It is of course the case that the four files accepted by parsers $A$, $C$, and $D$ are still accepted by both parsers $A$ and $C$, but this is not new information.)  Since parsers $A$ and $C$ only accept one new file when we exclude parser $D$, this suggests that the three parsers mostly agree on the \emph{de facto} format.
	
	On the other hand, parser $C$ accepts an additional $2$ files if we remove parser $A$ from the pair of parsers $\{A,C\}$.  This suggests that there is some disagreement between parsers $A$ and $C$, though it is perhaps somewhat minor.   In particular, since parser $C$ accepts files $4$ and $5$ in addition to the ones it accepts in common with parser $A$, it is likely that these two files contain some contentious aspect of the \emph{de facto} format, at least from parser $A$'s perspective.  Thus, when there are inconsistent subset relationships, we can identify which files are potentially at fault for this inconsistency.  To identify all files that are potentially contentious, it is important to realize that files represented in a given inconsistent subset relation may be cause for acceptance or rejection elsewhere in the weighted Dowker graph.  Specifically, the fact that files $4$ or $5$ are implicated in a contention between parsers $A$ and $C$ does not absolve these files from causing a contention with parser $D$ as well.
	
	We posit that 
	\begin{quote}
		the minimally acceptable \emph{de facto} format is likely embodied by a subgraph of the weighted Dowker graph which is an \emph{upwardly closed subset} and \emph{fully consistent}.  
	\end{quote}
	That is to say, each subset of parsers is in a consistent subset relation with each of its subsets.  For this example, such a subgraph consists of the following parser subsets $\{A,B\}$, $\{A,D\}$, $\{C,D\}$, $\{A\}$, $\{B\}$, $\{C\}$, and $\{D\}$.  Any file that is accepted by a subset of parsers \emph{other} than one of these is deemed an \emph{inconsistent file}, and potentially contains a contentious aspect of the \emph{de facto} format.  In this example, files $10$, $13$, $17, \dots, 20$ are inconsistent.

	\subsection{The \texttt{Govdocs1} Corpus \label{sec:Govdocs1}}
	
\begin{figure}[htpb]
		\centering
		\includegraphics[trim = 25mm 120mm 35mm 120mm, clip, width=\columnwidth,keepaspectratio]{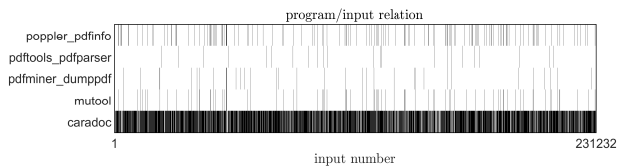}
		\caption{ \label{fig:Govdocs1Relation} The parser-file relation for the whole \texttt{Govdocs1} corpus, represented as a Boolean matrix. $1$'s (acceptances) are shown as white, $0$'s (rejections) as black.
		}
	\end{figure}

	To establish a baseline for parser agreement in a typical set of files, we processed the PDF portion of the \texttt{Govdocs1} corpus \cite{garfinkel2010digital} using five parsers (\texttt{caradoc} \cite{endignoux2016caradoc}, \texttt{mutool}, \texttt{pdfminer\_dumppdf}, \texttt{pdftools\-\_pdfparser}, and \texttt{poppler\_pdfinfo}). We ran these parsers to collect the output to \texttt{stderr} from each parser for analysis.
	\footnote{
		Although these parsers are open-source command line tools that are readily instrumented, using, e.g. \texttt{Cuckoo Sandbox} to monitor self-contained PDF software would allow us to go beyond (and also avoid) merely rendering files as in \cite{kuchta2018correctness}: besides collecting output to \texttt{stderr}, such an approach also enables the detection of nonevasive malware \cite{xu2016automatically}.
	}
	In particular, a parser sending \emph{any} output (resp., \emph{no} output) to \texttt{stderr} was deemed to \emph{reject} (resp., \emph{accept}) the file.  As in \S \ref{sec:toy} we summarized the result of this processing as a matrix, in which each row corresponds to a parser and each column corresponds to a file.  For visual clarity, we represent the matrix as the image shown in Figure \ref{fig:Govdocs1Relation}, in which acceptance and rejection are respectively indicated in white and black.
	
	\begin{figure}[htpb]
		\centering
		\includegraphics[trim = 25mm 120mm 35mm 120mm, clip, width=\columnwidth,keepaspectratio]{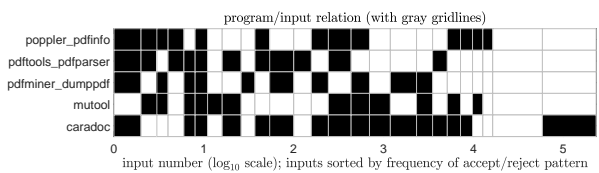}
		\caption{ \label{fig:Govdocs1RelationSorted} A sorted and scaled representation of the parser-file relation.
		}
	\end{figure}
	
	
	It is visually apparent from Figures \ref{fig:Govdocs1Relation} and \ref{fig:Govdocs1RelationSorted} that the \texttt{caradoc} parser rejects far more files than the others (this is a consequence of \texttt{caradoc}'s intended use as a PDF validator), but that there is wide agreement otherwise.  This intuition is borne out by actually estimating the probability that a typical file is accepted by a given parser, either individually or conditioned on acceptance by another parser. Using counts of acceptances/rejections over files in \texttt{Govdocs1} for the purpose of estimation, we have
	\begin{center}
		\begin{tabular}{|l|c|}
			\hline
			Parser & $\hat{\mathbb{P}}(\text{accept})$\\
			\hline
			\texttt{caradoc} & $0.2109$ \\
			\texttt{mutool} & $0.9691$ \\
			\texttt{pdfminer\_dumppdf} & $0.9882$ \\
			\texttt{pdftools\_pdfparser} & $0.9916$ \\
			\texttt{poppler\_pdfinfo} & $0.9489$\\
			\hline
		\end{tabular}
	\end{center}
	From this we see \texttt{caradoc} is the most stringent of the five parsers, accepting approximately one file out of every five.  The others are much more permissive, accepting most files.
	
	\begin{figure}[htpb]
		\centering
		\includegraphics[trim = 50mm 100mm 70mm 100mm, clip, width=.6\columnwidth,keepaspectratio]{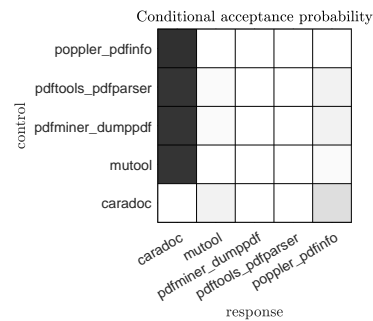}
		\caption{ \label{fig:Govdocs1Conditional} Pairwise comparison of parsers over the whole \texttt{Govdocs1} corpus. Greater conditional probabilities represented as brighter shades of grey.
		}
	\end{figure}
	
	The pairwise conditional probabilities, shown in Figure \ref{fig:Govdocs1Conditional}, are to be interpreted as giving the probability that the parser on the column label accepts a file given that the parser on the row has already accepted the file.  In virtue of the last row, one may conclude that if \texttt{caradoc} accepts a file, other parsers usually do too (again, this is a consequence of \texttt{caradoc}'s intended use as a PDF validator).  The middle block indicates that \texttt{mutool}, \texttt{pdfminer}, and \texttt{pdftools} usually agree.  The first column indicates that \texttt{caradoc}'s behavior cannot be inferred from the other parsers, aside from the fact that it usually rejects files that the others accept.
	
	\begin{figure}[htpb]
		\centering
		\includegraphics[trim = 55mm 105mm 50mm 105mm, clip, width=\columnwidth,keepaspectratio]{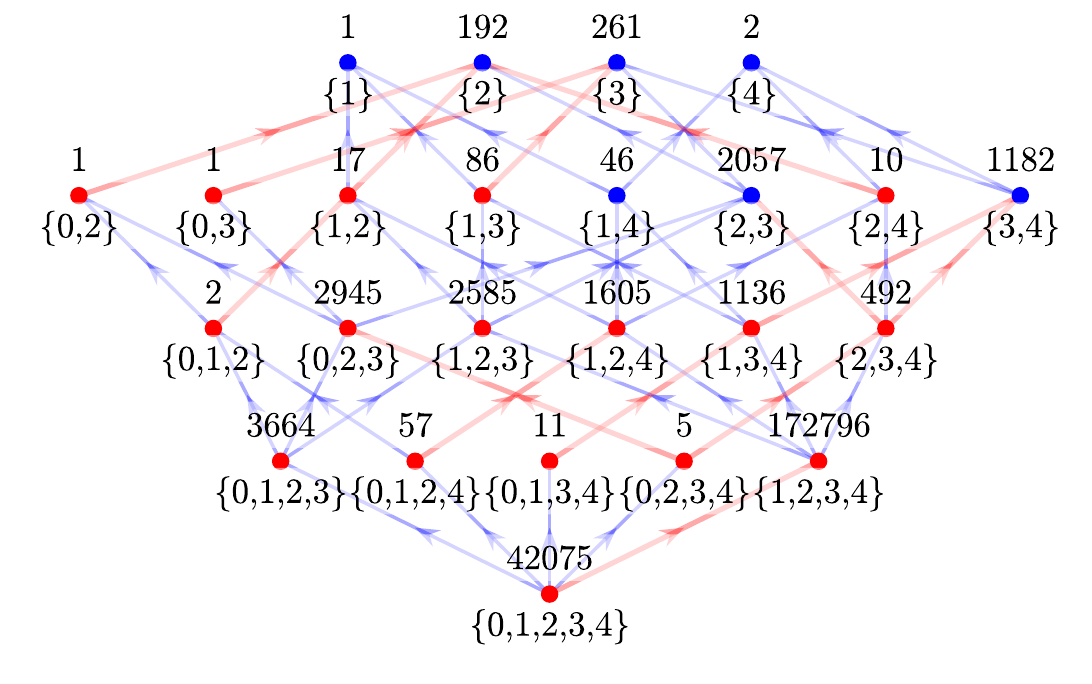}
		\caption{ \label{fig:Govdocs1Complex} Weighted Dowker graph for the \texttt{Govdocs1} corpus. Inconsistencies labeled in {\color{red}red}. Graph nodes correspond to the subsets of $\{0,\dots,4\}$ indicated below the nodes, where $0,\dots,4$ respectively correspond to \texttt{caradoc}, \texttt{mutool}, \texttt{pdfminer\_dumppdf}, \texttt{pdftools\_pdfparser}, and \texttt{poppler\_pdfinfo}. The corresponding weights are shown above the nodes. While this visual representation is more convenient than those in Figures \ref{fig:polyvenn5}, \ref{fig:Govdocs1Relation}, and \ref{fig:Govdocs1RelationSorted}, the algorithmic approach of \S \ref{sec:Theory} to identifying consistent subsets of the power set of programs is ultimately necessary.  
		}
	\end{figure}
	
	Although conditional probability is a useful tool, it is fundamentally pairwise.  Relationships among three or more parsers are not considered unless we lift our perspective to the power set of programs under consideration, which brings us to the entry point for TDT. With this in mind, we plot the weighted Dowker graph as before, in Figure \ref{fig:Govdocs1Complex}.
	Again, we can see (from the top row of the Dowker graph) that \texttt{pdfminer} (row $2$ in Figures \ref{fig:Govdocs1Relation} and \ref{fig:Govdocs1RelationSorted}, using zero-based indexing) and \texttt{pdftools} (row $3$) are (individually) the most permissive, since they accept many more files on their own that are otherwise not accepted.  The inconsistency at the bottom of the graph along the subset relation $(0,1,2,3,4) {\color{red}\to} (1,2,3,4)$ emphasizes \texttt{caradoc}'s stringency: even when all other parsers accept a file, \texttt{caradoc} rejects it more often than not.  Specifically, the number of files accepted by the set of all parsers is substantially smaller than the number of files accepted by the set of parsers not including \texttt{caradoc}.
	
	To identify which files are potentially causing the inconsistency -- and thus could be representative of differences in the \emph{ad hoc} format implemented by different parsers -- we trace back to the columns of the matrix shown in Figure \ref{fig:Govdocs1Relation} to identify the inconsistent files.  Specifically, we identify those files corresponding to subsets of parsers that are not contained in the upwardly closed, consistent subgraph of the weighted Dowker graph.  Under this definition of an inconsistent file, it happens that \emph{every file} is a potential cause of inconsistency \emph{somewhere} in the weighted Dowker graph
	\footnote{
		The identity of the files is not readily recovered from Figure \ref{fig:Govdocs1Complex} as it requires identifying columns in Figure \ref{fig:Govdocs1Relation}.
	} 
	shown in Figure \ref{fig:Govdocs1Complex}.  Even though a given edge in the graph may be consistent -- and the files accepted by the parsers involved -- those same files happen to be inconsistent at some other edge.  To provide more useful information, we relax the notion of an \emph{inconsistent file} to an integer \emph{inconsistency score} for each file.  To assign this score, we iterate over subsets of parsers, restricting our attention to just those parsers for the computation of the inconsistent files.  The \emph{inconsistency score} is the number of subsets of parsers that deem the file to be inconsistent.
	
	The inconsistency score is more tolerant of differences in parser behavior.  For instance, every file is inconsistent for the set of all five parsers, but most files are not deemed inconsistent when \texttt{caradoc} is excluded from consideration.  On the basis of statistics alone, it would seem wise to exclude \texttt{caradoc} entirely, but in fact some subsets of parsers that include \texttt{caradoc} do not deem every file inconsistent.  An even-handed approach, as we suggest here, is therefore warranted.
	
	\begin{figure}
		\centering
		\includegraphics[trim = 25mm 120mm 35mm 120mm, clip, width=\columnwidth,keepaspectratio]{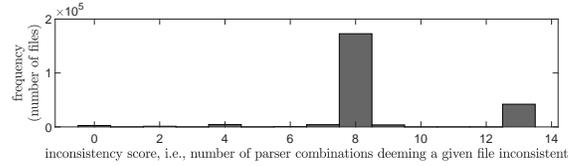}
		\caption{ \label{fig:Govdocs1Scores} Histogram of inconsistency scores for the \texttt{Govdocs1} corpus. 
		}
	\end{figure}
	
	Figure \ref{fig:Govdocs1Scores} shows a histogram of inconsistency scores for all files.  Intuitively, binning files by their inconsistency score appears to cluster files based on ``dialects'' of the \emph{de facto} format.  We also see a clear boundary in the distribution: most files are deemed inconsistent by $7$ or more subsets of parsers (\texttt{caradoc}'s behavior results in most of these), but few files have less than that. These few files are worth investigating further.
	
	\begin{figure}
		\centering
		\includegraphics[trim = 25mm 120mm 35mm 120mm, clip, width=\columnwidth,keepaspectratio]{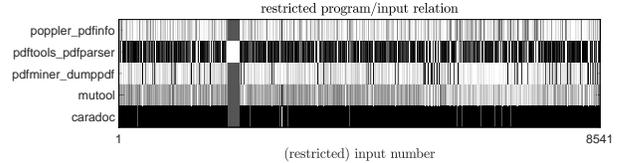}
		\caption{ \label{fig:Govdocs1Restricted} The \texttt{Govdocs1} parser-file relation restricted to files with inconsistency score less than $7$.
		}
	\end{figure}
	
	As a first pass, we can restrict the relation shown in Figure \ref{fig:Govdocs1Relation} to only the files with inconsistency score less than $7$. We plot this relation matrix in Figure \ref{fig:Govdocs1Restricted}, and see some striking features. Most of these files are unusual because only \texttt{pdfminer} and \texttt{pdftools} accepted them. Since these parsers are the most permissive, the files in question are probably on the edge of acceptability, and are quite possibly out of spec altogether.
	
	A clear subset of files is accepted only by \texttt{pdftools}.  This subset contained $261$ files.  To explore the causes of these files being selected as unusually inconsistent, we examined the \texttt{stderr} output from the parsers that rejected these files.  We found that $212$ of these files were actually being accepted, even though there was \texttt{stderr} output.  Of the remaining $37$ files, one of them was not a PDF file at all, six of them crashed one of the parsers, and the remainder had malformed objects.  Although the purpose of our approach is not to diagnose specific format violations, the similarity of the kinds of \texttt{stderr} reports indicates that inconsistency score does appear to cluster contentious files with similar problems.

	\subsection{A test corpus \label{sec:Hackathon}}
	
	We also applied TDT to a test corpus of 120 files that were nominally compliant with the PDF standard, and most of which were designed to violate precepts of the format.  The test corpus was designed by an independent performer to evaluate the ability of performers on the DARPA SafeDocs program to identify format violations during a hackathon, using \texttt{Govdocs1} in advance to produce a \emph{de facto definition} of what constitutes a valid PDF file.  Of these 120 files, 82 were deemed to be non-compliant.
	
	Unsurprisingly, the test corpus behaved substantially differently than \texttt{Govdocs1}, and in particular was collectively much less adherent to any \emph{de facto} standard: 
	\begin{center}
		\begin{tabular}{|l|c|}
			\hline
			Parser & $\hat{\mathbb{P}}(\text{accept})$\\
			\hline
			\texttt{caradoc} & $0.1417$ \\
			\texttt{mutool} & $0.8417$ \\
			\texttt{pdfminer\_dumppdf} & $0.6500$ \\
			\texttt{pdftools\_pdfparser} & $0.7333$ \\ 
			\texttt{poppler\_pdfinfo} & $0.3583$\\
			\hline
		\end{tabular}
	\end{center}
	
	\begin{figure}
		\centering
		\includegraphics[trim = 25mm 120mm 35mm 120mm, clip, width=\columnwidth,keepaspectratio]{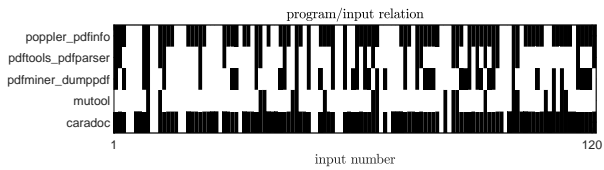}
		\caption{ \label{hackathonRelation} The hackathon parser-file relation.
		}
	\end{figure}
	
	\begin{figure}
		\centering
		\includegraphics[trim = 25mm 120mm 35mm 120mm, clip, width=\columnwidth,keepaspectratio]{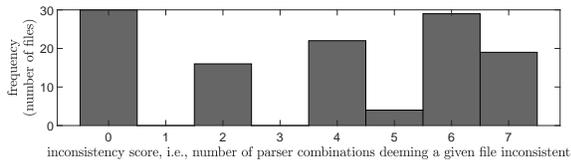}
		\caption{ \label{fig:hackathonScores} Histogram of inconsistency scores for the hackathon dataset. 
		}
	\end{figure}
	
	We analyzed this corpus both in its own right and using a binary PDF validity classifier ``trained'' by using TDT on \texttt{Govdocs1} as in \S \ref{sec:Govdocs1}.  Because of our experience on the \texttt{Govdocs1} data, we decided to exclude \texttt{pdftools} from the analysis, but retained the other four parsers, namely \texttt{caradoc}, \texttt{mutool}, \texttt{pdfminer}, and \texttt{poppler}.  The binary classifier selected a file as being not compliant with the \emph{de facto} standard whenever two or more parsers rejected it.  (Note that for the purposes of this exercise, actual compliance with the \emph{de facto} standard itself was adjudicated by the independent performer as noted above.)  Since \texttt{caradoc} was again the most stringent parser, this usually meant that files marked as non compliant by the binary classifier were rejected by \texttt{caradoc} and another parser.
	
	Of the 120 files, the binary classifier's precision was 92\% and its recall was 97\%, resulting in an F1 score of 95\%. 
	
	To delve into the reasons why specific files were rejected by the parsers according to our binary classifier, we computed the distribution of inconsistency scores for the test corpus (using the same parsers as in \S \ref{sec:Govdocs1}), shown in Figure \ref{fig:hackathonScores}. Note that this histogram is much flatter than the one for \texttt{Govdocs1}, which is probably an artifact of the strategy for causing file malformations (note also that far fewer files are present than for \texttt{Govdocs1}).
	
	Nearly all of the files selected by our binary classifier had an inconsistency score of less than $3$ or equal to $6$.  Indeed, using that as an \emph{ex post facto} decision rule results in a precision of 100\%, recall of 85\%, and an F1 score of 91.5\%.
	
	Below, we describe the results from investigating the spikes visible at inconsistency scores $2$, $4$, and larger than $5$ and explore the reasons for parse failures that occurred.  Files with inconsistency score $0$ were all non compliant files, which essentially means that all of the parsers agreed about the disposition of these files.

	\paragraph{Inconsistency Score $2$}
	
	The files with inconsistency score $2$ happened to be those files that were accepted by \texttt{mutool} (and ultimately \texttt{pdftools} also, though we did not use its output) but rejected by the other parsers.  All of these files were marked by our binary classifier as being non-compliant, and all of these files were indeed not compliant.
	
\begin{figure}
		\centering
		\includegraphics[trim = 25mm 100mm 35mm 105mm, clip, width=\columnwidth,keepaspectratio]{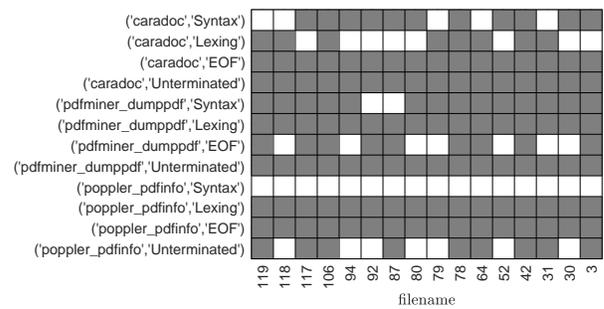}
		\caption{ \label{fig:hackathonError2} Table of errors parsers send to \texttt{stderr} for files with inconsistency scores of $2$. White signifies the presence of a keyword match; gray signifies its absence.
		}
	\end{figure}

	There are numerous different error reports produced on \texttt{stderr} for the parsers that rejected each file with inconsistency score $2$.  To quantify what kinds of error messages were produced on \texttt{stderr}, we performed a simple keyword search.  These result of this search is summarized in Figure \ref{fig:hackathonError2}; a white box indicated that a particular keyword search term was found in that parser's \texttt{stderr} (column) in a given file (row).  Note that the keywords used for the search are different for each parser, since they structure their output differently.  However, each file (row) contains a match for at least one of the keywords given for each parser.  This suggests that we have a complete, if coarse, characterization of the errors produced. \footnote{See also the error message clustering of \cite{kuchta2018correctness}.}

	Given the results of the keyword search in Figure \ref{fig:hackathonError2}, it appears that these files mostly contain syntax and lexing issues due to files ending in unexpected places.  Depending on exactly where the files end, different parsers appear to handle the file differently.  For instance, \texttt{caradoc} claimed that the files contain syntax or lexing errors, and terminated immediately at that point.  As a result, \texttt{caradoc} never noticed that these files seem to end oddly, because it had already aborted processing.  In contrast, the other parsers did not notice the error as aggressively, but instead continued processing.  They only noticed that something was amiss when they hit the end of the file unexpectedly.
	
	\paragraph{Inconsistency Score $4$}
	
	There were three different classes of behavior for files with inconsistency score $4$:
	\begin{itemize}
		\item 17 files were accepted by all parsers
		\item 2 files were rejected by \texttt{caradoc}, \texttt{mutool}, and \texttt{poppler}, and
		\item 3 files were rejected by \texttt{caradoc} and \texttt{pdfminer}.
	\end{itemize}
	The binary classifier selected the 5 files in the latter two categories as not compliant.  Of all of the inconsistency score $4$ files, the binary classifier made exactly one error: one file in the latter category (rejected by both \texttt{caradoc} and \texttt{pdfminer}) was actually deemed to be compliant.  
	
	\paragraph{Highly inconsistent files}
	
\begin{figure}
		\centering
		\includegraphics[trim = 25mm 120mm 35mm 120mm, clip, width=\columnwidth,keepaspectratio]{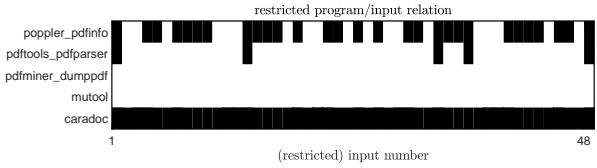}
		\caption{ \label{hackathonRestricted} The hackathon parser-file relation restricted to files with inconsistency scores greater than $5$.
		}
	\end{figure}

	We can also restrict our attention to only the files with high inconsistency scores.  These files should be expected to exhibit features that the parsers will generally find contentious -- not specifically files that are badly formatted, but formatted in a way that at least some parsers find objectionable while others do not.  If we restrict our attention to the relation between parsers and files with inconsistency score greater than $5$, the result is shown in Figure \ref{hackathonRestricted}. These highly inconsistent files are generally accepted by most parsers, but definitely not \texttt{caradoc} and usually not \texttt{poppler}. These files likely have only have small malformations. The error messages reported on \texttt{stderr} are highly variable but mostly seem to be syntax issues. It might be the case that data payloads (which the parsers don't read directly) are corrupted, or are of incorrect length.
	
	\section{Theoretical Details \label{sec:Theory}}

	\subsection{\label{sec:Weighted}Weighted Venn diagrams}
	For $m, n \in \mathbb{N}$, consider a set of $m$ programs and a set of $n$ inputs. Without loss of generality, take these sets to be $[m] := \{1,\dots,m\}$ and $[n]$, respectively. Let $R$ be a relation between $[m]$ and $[n]$ that models the behavior of programs on inputs \emph{\`a la} $(j,k) \in R$ if and only if program $j$ accepts input $k$. We also abusively write $R \in M_{m,n}(\mathbb{F}_2)$ or $R \in M_{m,n}(\mathbb{N})$ as circumstances and convenience dictate.
	
\begin{figure}[htpb]
		\centering
		\includegraphics[trim = 95mm 130mm 95mm 125mm, clip, scale=0.75,keepaspectratio]{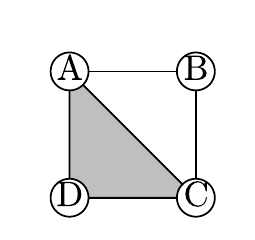}
		\includegraphics[trim = 95mm 130mm 95mm 125mm, clip, scale=0.75,keepaspectratio]{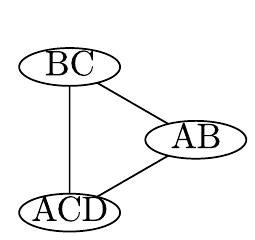}
		\includegraphics[trim = 95mm 130mm 95mm 125mm, clip, scale=0.75,keepaspectratio]{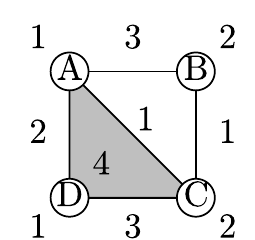}
		\includegraphics[trim = 95mm 130mm 95mm 125mm, clip, scale=0.75,keepaspectratio]{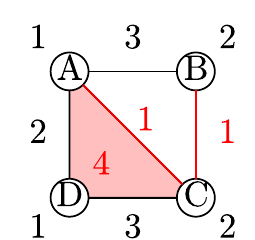}
		\caption{ \label{fig:Dowker} (Left) The Dowker complex associated to our toy example of \S \ref{sec:toy}. The ``hole'' (which as it happens can be detected using simplicial homology \cite{ghrist2014elementary}) suggests that program $B$ behaves differently than the other programs. (Center left) The Dowker complex of the dual/transposed relation is homotopically equivalent: since it has 10 vertices (corresponding to unique matrix columns), we show yet another homotopically equivalent (collapsed) complex with vertices corresponding to non-redundant columns. (Center right) The weighted Dowker complex associated to our toy example. (Right) The six inputs corresponding to the red faces (i.e., the complement of the interior in the Alexandrov topology) cause inconsistent program behaviors.
		}
	\end{figure}

	To any relation $R$, we can associate a \emph{abstract simplicial complex}, i.e., a hypergraph with all sub-hyperedges. Each hyperedge is called a \emph{face} (or \emph{simplex}), and maximal faces are called \emph{facets}. The construction is straightforward: vertices correspond to elements of $[m]$, i.e., rows of $R$ or programs; faces correspond to subsets of $[n]$ represented by columns of $R$, i.e., sets of programs that accept a common input.  Specifically, the \emph{Dowker complex $D(R)$ for a relation $R$} has $[m]$ as its set of vertices, and $[p_0,p_1, \dotsc, p_k]$ is a simplex of $D(R)$ whenever there is a $f \in [n]$ such that $(p_i,f)\in R$ for some $i=0, \dotsc, k$.  By a theorem of Dowker \cite{Dowker1952,ghrist2014elementary}, the two abstract simplicial complexes respectively formed in this way from $R$ and its transpose are topologically equivalent, and hence it is only a minor abuse to call either of these \emph{the Dowker complex} associated to $R$ (Figure \ref{fig:Dowker}).
	
	In practice, $m \ll n$ (i.e., we have many more inputs than programs), and so the Dowker complex associated to $R$ is not very informative (there are either too many redundant simplices or vertices, depending on which convention we choose for the Dowker complex). But we can ``project'' $R$ onto the power set $2^{[m]}$ as follows. First, define a matrix $P \in M_{m,2^m}(\mathbb{F}_2)$ so that the $\ell$th column of $P$ is the binary representation of $\ell-1$, i.e. $\ell = 1+\sum_{j = 1}^m P_{j \ell} 2^{j-1}$. Next, we define a weight tuple $w \in \mathbb{N}^{2^m}$ by counting the number of columns of $R$ that equal a given column of $P$. That is, $w_\ell := |\{k : R_{\cdot,k} = P_{\cdot,\ell}\}|$.  
	
\begin{figure}[htpb]
		\centering
		\includegraphics[trim = 80mm 114mm 80mm 114mm, clip, scale=0.6,keepaspectratio]{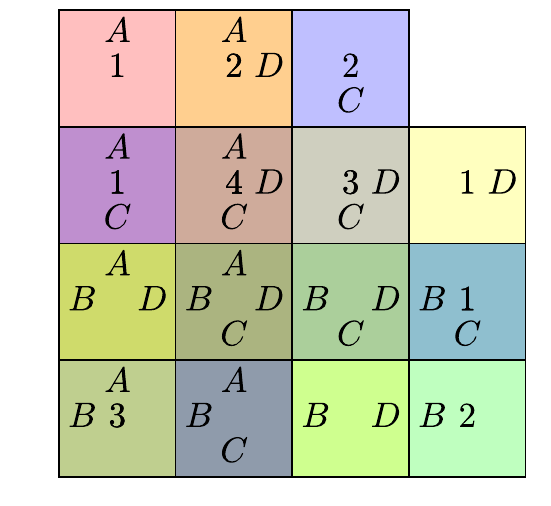}
		\caption{ \label{fig:Venn4} The weighted Venn diagram for our toy example of \S \ref{sec:toy}.
		}
	\end{figure}

	For each subset of programs under consideration, the associated weight is the number of files that (exactly) that subset of programs accepts.  Accordingly, we call the weight tuple $w$ the \emph{weighted Venn diagram of $R$}, or often simply the \emph{diagram} (Figure \ref{fig:Venn4}).
	By attaching weights to faces, we obtain a \emph{weighted Dowker complex} (Figure \ref{fig:Dowker}) and an equivalent weighted partial order whose representation as a digraph is called the \emph{weighted Dowker graph}, and which we have already encountered (Figure \ref{fig:toyComplex}).
	
	In this way, the weights can also be thought of as a function $w: D(R) \to \mathbb{N}$ on the Dowker complex defined by
	\begin{equation*}
		w(\sigma) := \#\{ f\in [n] : (p \in \sigma\text{ and } (p,f)\in R) \text{ or } (p \notin \sigma \text{ and } (p,f) \notin R) \}
	\end{equation*}
	for each simplex $\sigma \in D(R)$.
	
	\subsection{\label{sec:Consistency}Deficiency and consistency}
	We should hope that programs agree more than they disagree, so that we can trust their agreement in which files are acceptable. To this end, we call a region of a diagram (i.e., a subset of programs $\subseteq [m]$) \emph{deficient} if its weight is less than the weight of any of its facets (regions which correspond to subsets of the set of programs for the region in question, where the subsets have one fewer element than the original). Intuitively, a deficient region represents a set of programs that collectively disagree with another program. 
	
	We call a diagram \emph{inconsistent} if any of its regions are deficient. Otherwise, a diagram is \emph{consistent}.  Rather simply, the weight function $w$ is consistent whenever it is an order preserving function.  If $w$ is not consistent, say for two simplices $\sigma \subseteq \tau$, we have that $w(\sigma) > w(\tau)$, then a given file $f \in [n]$ is \emph{inconsistent for the pair $\sigma,\tau$} if there is a $p \in \tau \backslash \sigma$ such that $(p,f) \notin R$.  If $w(\sigma) \le w(\tau)$, then no files are deemed inconsistent for that pair of simplices.
	
	\subsection{\label{sec:Basics}Basic results}
	
	\textsc{Definition.} A \emph{filtration} of an abstract simplicial complex is a reverse inclusion-ordered family of subcomplexes.
	
	\
	
	\textsc{Lemma.} The diagram of $R$ is consistent if and only if its weights produce a filtration on the abstract simplicial complex given by the subsets of the domain of $R$. 
	
	\
	
	\textsc{Proof.} For weights to produce a filtration on a simplicial complex, the weight assigned to any simplex must be no less than the weight assigned to any simplex in the its boundary. If a diagram is inconsistent, then its deficient region corresponds to a simplex whose filtration value is too small. If a diagram's weights cannot produce a filtration, then some simplex must have a filtration value smaller than the value of one of the simplices in its boundary. A simple induction shows that some simplex between these two (in the boundary of the former and the coboundary of the latter, inclusive) corresponds to a region of the diagram that must be deficient.
	\hfill $\Box$
	
	\
	
	\textsc{Definition.} The diagram on $R'$ is a \emph{subdiagram} of the diagram on $R$ if and only if $R'$ is a domain restriction of $R$.
	
	\
	
	\textsc{Lemma.} If a diagram is consistent, then all of its subdiagrams are consistent.
	
	\
	
	\textsc{Proof.} Let the diagram of $R$ have weights $w$. By induction, we only have to prove the subdiagram obtained by removing one program to be consistent. Let $\nabla_j R$ be the restriction of $R$ obtained by removing the $j$th program. The weights $\nabla_j w$ of the diagram of $\nabla_j R$ are simply $\nabla_j w(X) = w(X) + w(X \cup \{j\})$ for $X \subseteq [m] \backslash \{j\}$. To prove that any region $X$ in the subdiagram is not deficient, we cite that both of the regions $X$ and $X \cup \{j\}$ in the original diagram are not deficient, and that inequalities are additive. The result follows.
	\hfill $\Box$
	
	\
	
	\textsc{Remark.} If a diagram is inconsistent, then any diagram for which it is a subdiagram is also inconsistent. It is simple to check whether the diagram of a relation with only one program is consistent.
	
	\
	
	\textsc{Remark.} If an inconsistent diagram has $N$ deficient regions, then all of its diagrams have at most $N$ deficient regions, usually fewer.

	\subsection{\label{sec:Algorithm}Algorithm}
	
	The results in the previous section suggest the following algorithm to, given a diagram, produce a consistent subdiagram:
	
	\
	
	We are given the diagram of some relation $R$. We start by restricting $R$ to only those programs for which the associated diagram is consistent --- call this restriction $\nabla_\forall R$. Then, we iteratively remove programs until the remaining relation produces a consistent diagram:
	
	Let the diagram of $\nabla_\forall R$ have deficient regions $d _1, \dots, d _N$. Let $d _i$ be the region with the greatest number of programs (if there are multiple such regions, ties are broken by deficiency, i.e. the difference between a region's weight and its faces' largest weight; further ties are broken arbitrarily). Consider the face of $d _i$ with the greatest weight (ties broken arbitrarily): this face includes all but one of the programs in $d _i$ --- call this excluded program the $j$th one. We restrict $\nabla_\forall R$ to all programs in its domain except $j$, create the subdiagram, and repeat.
	
	\
	
	We have not yet analyzed the computational complexity of this algorithm, not least since in practice the number of programs should usually be small enough for exponential-time algorithms to run in reasonable time. We also note that this algorithm is not guaranteed to produce a maximal restricted relation, i.e., a relation to which no program could be added without resulting in an inconsistent diagram. We also note that the algorithm's result is deterministic if and only if ties are broken deterministically.

	\subsection{\label{sec:SimplestExample}An even simpler example}
	
	Consider
	\setcounter{MaxMatrixCols}{20}
	\[R = \left(\begin{smallmatrix}
	0 & 0 & 1 & 1 & 1 & 0 & 0 & 0 & 1 & 0 & 1 & 1 & 1 & 0 \\ 
	0 & 0 & 0 & 0 & 1 & 0 & 1 & 1 & 0 & 1 & 0 & 1 & 0 & 1 \\
	1 & 1 & 0 & 1 & 0 & 0 & 0 & 0 & 1 & 1 & 0 & 1 & 1 & 0
	\end{smallmatrix}\right).\]
	Then (see Figure \ref{fig:venn}) $w = (1,2,3,1,2,3,1,1)$ and the region corresponding to $\{A, B, C\}$ is deficient (by its face $\{A, C\}$), as well as the regions $\{A, B\}$ and $\{B, C\}$ (by any of their faces). If we construct the subdiagrams of singleton restrictions of its relation (see Figure \ref{fig:singleton}), we see we should exclude program $B$. The diagram of the resulting relation (left panel of Figure \ref{fig:vennFinal}) has two deficient regions $\{A\}, \{C\}$: we arbitrarily choose $\{C\}$. This region has only one face ($\varnothing$), and this face excludes $C$, so we restrict the relation to exclude $C$. The resulting relation has a consistent diagram (right panel of Figure \ref{fig:vennFinal}).
	
	\def\firstcircle{(60:1cm) circle (1cm)}
	\def\secondcircle{(0,0) circle (1cm)}
	\def\thirdcircle{(0:1cm) circle (1cm)}
	\begin{figure}[htbp]
		\centering
		\begin{tikzpicture}[scale = 1,every node/.style={transform shape}]
		\begin{scope}[fill opacity=0.25]
		\fill[red] \firstcircle;
		\fill[green] \secondcircle;
		\fill[blue] \thirdcircle;
		\end{scope}
		\draw \firstcircle;
		\draw \secondcircle;
		\draw \thirdcircle;
		\node (v000) at (-.75,1.25) {$1$};
		\node (v100) at (.5,1.25) {$2$};
		\node (v010) at (-.5,-.25) {$3$};
		\node (v110) at (-.1,.6) {$1$};
		\node (v001) at (1.5,-.25) {$2$};
		\node (v101) at (1.1,.6) {$3$};
		\node (v011) at (.5,-.5) {$1$};
		\node (v111) at (.5,.125) {$1$};
		\node (n1) at (.5,1.6) {{\color{red} $A$}};
		\node (n2) at (-.5,-.6) {{\color{green} $B$}};
		\node (n3) at (1.5,-.6) {{\color{blue} $C$}};
		\end{tikzpicture}
		\includegraphics[trim = 80mm 125mm 80mm 125mm, clip, scale=0.75,keepaspectratio]{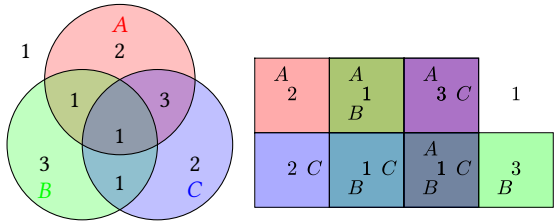}
		\caption{ \label{fig:venn} (Left) Venn diagram for the projected relation. There is a ``gap'' between {\color{green}the set of 3 files accepted by program $B$ and rejected by programs $A$ and $C$} and the sets of {\color{red}2 files accepted by program $A$ and rejected by programs $B$ and $C$}, of {\color{purple}3 files accepted by programs $A$ and $C$ and rejected by program $B$}, and of {\color{blue}2 files accepted by program $C$ and rejected by programs $A$ and $B$}. (Right) Another representation of the same diagram as a $(1,2)$ polyVenn.
		}
	\end{figure}

	\begin{figure}[htbp]
		\centering
		\begin{tikzpicture}[scale = 1,every node/.style={transform shape}]
		\begin{scope}[fill opacity=0.25]
		\fill[red] (-3,0) circle (1);
		\fill[green] (0,0) circle (1);
		\fill[blue] (3,0) circle (1);
		\end{scope}
		\draw (-3,0) circle (1);
		\draw (0,0) circle (1);
		\draw (3,0) circle (1);
		\node (v1) at (-3.5,.5) {$7$};
		\node (v-1) at (-4,1) {$7$};
		\node (v2) at (-.5,.5) {$6$};
		\node (v-2) at (-1,1) {$8$};
		\node (v3) at (2.5,.5) {$7$};
		\node (v-3) at (2,1) {$7$};
		\node (n1) at (-3,0) {{\color{red} $A$}};
		\node (n2) at (0,0) {{\color{green} $B$}};
		\node (n3) at (3,0) {{\color{blue} $C$}};
		\end{tikzpicture}
		\caption{ \label{fig:singleton} Diagrams of the singleton restrictions of the above relation. $B$ should be excluded, since its diagram is inconsistent. 
		}
	\end{figure}
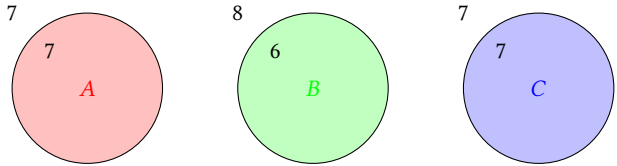

	\begin{figure}[htbp]
		\centering
		\begin{tikzpicture}[scale = 1,every node/.style={transform shape}]
		\begin{scope}[fill opacity=0.25]
		\fill[red] \firstcircle;
		\fill[blue] \thirdcircle;
		\end{scope}
		\draw \firstcircle;
		\draw \thirdcircle;
		\node (v000) at (-.75,1.25) {$4$};
		\node (v100) at (.5,1.25) {$3$};
		\node (v001) at (1.5,-.25) {$3$};
		\node (v101) at (1.1,.6) {$4$};
		\node (n1) at (.5,1.6) {{\color{red} $A$}};
		\node (n3) at (1.5,-.6) {{\color{blue} $C$}};
		\end{tikzpicture}
		\quad \quad
		\begin{tikzpicture}[scale = 1,every node/.style={transform shape}]
		\begin{scope}[fill opacity=0.25]
		\fill[red] \firstcircle;
		\end{scope}
		\draw \firstcircle;
		\node (v000) at (-.75,1.25) {$7$};
		\node (v100) at (.5,1.25) {$7$};
		\node (n1) at (.5,1.6) {{\color{red} $A$}};
		\end{tikzpicture}
		\caption{ \label{fig:vennFinal} (Left) Diagram of the above relation's restriction to $\{A, C\}$. This diagram has two deficient faces. (Right) Diagram of the above relation's restriction to $\{A\}$. This diagram is consistent.
		}
	\end{figure}
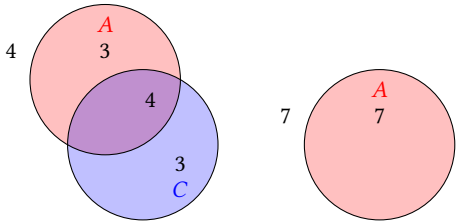

	\subsection{\label{sec:Sheaf}Sheaf interpretation}
	
	(NB. This section assumes familiarity with the notion of a \emph{sheaf}, for which see, e.g. \cite{ghrist2014elementary}.) 
	
	The algorithm of \S \ref{sec:Algorithm} is similar to the algorithm presented in \cite{Praggastis2016} (see also \cite{Robinson2015}), which deterministically yields all maximal sections of a sheaf (albeit at higher computational cost). We can convert our question to be suitable for this algorithm:
	
	Let $\mathcal{X}$ be the abstract simplicial complex of sets of programs. We give this space the ``reverse'' Alexandrov topology (i.e. we switch the usual definitions of upper and lower sets). Let $\mathcal{S}$ be a sheaf over $\mathcal{X}$ defined as follows. The sections of $\mathcal{S}$ for a $k$-simplex are partitions of the set of files into $2 ^{k + 1}$ possibly empty classes (i.e., partitions into the regions of a Venn diagram of the programs comprising that simplex). The restriction maps of $\mathcal{S}$ are coarsenings of these partitions: the section of a $k$-simplex should have each class equal to the union of the corresponding $2 ^{j - k}$ classes in the section of any $j$-simplex on its coboundary. We also define \emph{consistency functions} \cite{robinson2018assignments,robinson2019hunting} as follows: an assignment is consistent at some simplex if all the restriction maps produce the same section, and if that section corresponds to a consistent diagram (as discussed above).
	
	\begin{figure}[htbp]
		\centering
		\begin{tikzpicture}[scale = 1,every node/.style={transform shape}]
		\begin{scope}[fill opacity=0.25]
		\fill[red] \firstcircle;
		\fill[green] \secondcircle;
		\fill[blue] \thirdcircle;
		\end{scope}
		\draw \firstcircle;
		\draw \secondcircle;
		\draw \thirdcircle;
		\node (v000) at (-.75,1.25) {$1$};
		\node (v100) at (.5,1.25) {$2$};
		\node (v010) at (-.5,-.25) {$3$};
		\node (v110) at (-.1,.6) {$20$};
		\node (v001) at (1.5,-.25) {$4$};
		\node (v101) at (1.1,.6) {$30$};
		\node (v011) at (.5,-.5) {$40$};
		\node (v111) at (.5,.125) {$900$};
		\node (n1) at (.5,1.6) {{\color{red} $A$}};
		\node (n2) at (-.5,-.6) {{\color{green} $B$}};
		\node (n3) at (1.5,-.6) {{\color{blue} $C$}};
		\end{tikzpicture}
		\caption{ \label{fig:venn2} Weighted Venn diagram for a projected relation. 
		}
	\end{figure}
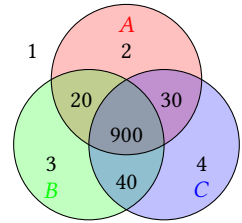
	
	\begin{figure}[htbp]
		\centering
		\begin{tikzpicture}[->,>=stealth',shorten >=1pt,scale = 0.85,every node/.style={transform shape}]]
		\node [draw, align=center] (vA) at (-3.5,3) {$\{A\}$; $\mathbb{N}^4$; \\ $(2,20,30,900)$};
		\node [draw, align=center] (vB) at (0,3) {$\{B\}$; $\mathbb{N}^4$; \\ $(3,20,40,900)$};
		\node [draw, align=center] (vC) at (3.5,3) {$\{C\}$; $\mathbb{N}^4$; \\ $(4,30,40,900)$};
		\node [draw, align=center] (vAB) at (-3.5,1.5) {$\{A,B\}$; $\mathbb{N}^6$; \\ $(2,3,30,40,20,900)$};
		\node [draw, align=center] (vAC) at (0,1.5) {$\{A,C\}$; $\mathbb{N}^6$; \\ $(2,4,20,40,30,900)$};
		\node [draw, align=center] (vBC) at (3.5,1.5) {$\{B,C\}$; $\mathbb{N}^6$; \\ $(3,4,20,30,40,900)$};
		\node [draw, align=center] (vABC) at (0,0) {$\{A,B,C\}$; $\mathbb{N}^7$; \\ $(2,3,4,20,30,40,900)$};
		\draw (vABC) to (vAB);
		\draw (vABC) to (vAC);
		\draw (vABC) to (vBC);
		\draw (vAB) to (vA);
		\draw (vAB) to (vB);
		\draw (vAC) to (vA);
		\draw (vAC) to (vC);
		\draw (vBC) to (vB);
		\draw (vBC) to (vC);
		\end{tikzpicture}
		\caption{ \label{fig:sheaf_small} ``Reverse'' face order ($\{\bullet\}$); sheaf ($\mathbb{N}^\bullet$); global section ($\in \mathbb{N}^\bullet$) for the example shown in Figure \ref{fig:venn2}. Components in the stalks are ordered lexicographically.
		}
	\end{figure}
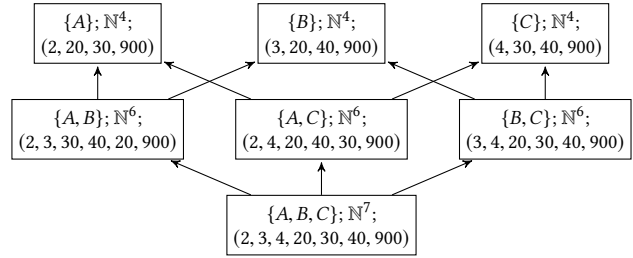
	
	The example in Figure \ref{fig:venn2} is shown as a global section of a sheaf in Figure \ref{fig:sheaf_small}. Instead of listing the files themselves, this particular sheaf merely retains the counts of files in each region.  Although not shown, the components in the stalks are ordered lexicographically.  The maximal elements in the partial order still have rather elaborate sections.  This indicates that the sheaf really ``wants'' to be built on the \emph{intersection lattice} of the sets $\{A,B,C\}$ rather than on $\mathcal{X}$ as we have done. We leave for future work a proof that our construction is sufficient (and perhaps minimally sufficient) to reconstruct the sheaf over the intersection lattice.  
	
	Summarizing, the present sheaf stores the partition of the files into all combinations of programs at the highest simplex, and stores simplifications of this partition (i.e. forgetting some programs) in the other simplices. The restriction maps simply ensure that the ``simplified'' partitions are properly defined. The consistency functions' primary purpose, rather than ensuring the globality of an assignment, is to check the consistency of Venn diagrams (the whole sheaf formulation is just a way to apply the algorithm of \cite{Praggastis2016} to this notion of consistency). From the perspective of implementations, it is convenient to  define sections in terms of the number of files in each partition class, as in Figure \ref{fig:sheaf_small}.

	\subsection{\label{sec:Selection}Selection of files}
	
	Once we run the algorithm from \S\ref{sec:Algorithm}, the weights indeed yield a filtered simplicial complex. Now an obvious strategy is to pick a suitable filtration value and exclude files corresponding to simplices that appear at lower filtration values.
	
	In the example of figure \ref{fig:venn2}, choosing a threshold of 20 would eliminate the 10 (of 1000 total) files that are accepted by at most one program; a threshold of 900 would eliminate the 100 files that are accepted by at most two programs. There is not any obvious reason to consider another choice unless there is nontrivial connectivity (an obvious criterion is to conservatively preclude that possibility, at least using homology in dimension zero, i.e., to require that there be a single connected component). But suppose now that the numbers 40 and 900 were replaced by 240 and 700, respectively: then another natural choice emerges.

	\section{Remarks \label{sec:Remarks}}
	
	
	
	Relations are ubiquitous, and it is likely that (the mathematical ideas behind) TDT can be applied more generally. We outline a progressive widening of scope here:
	
	\begin{itemize}
		\item Using delta debugging \cite{zeller2002simplifying,hodovan2017coarse} to provide inputs to TDT could help patch diverse implementations.
		\item The technique of \cite{sivakorn2017hvlearn} learns deterministic finite automata (or equivalently, regular expressions) corresponding to SSL/TLS common names on a per-implementation basis. Since DFAs/ regexes are closed under union and intersection, TDT could extend this technique to determine a \emph{de facto} standard for such common names by considering multiple implementations.
		\item Coupling TDT with differential fuzzing of machine learning or data science algorithms with similar functionality \cite{pei2017deepxplore,guo2018dlfuzz,srisakaokul2018multiple,patterson2018teaching} is a particularly fertile ground for future work.
		\item TDT should also be applicable to binary classifers more generally \cite{jaffe2015estimating}, though even our specific context of parser/format evaluation is adjacent to applications such as data structure evaluation \cite{molina2019training} and malware multiscanning \cite{sakib2020maximizing}.
	\end{itemize}

	\subsection{A useful relation ``product'' \label{sec:product}}
	
	Let $R \subseteq [m] \times [n]$ be a relation between programs and inputs as usual, and let $S \subseteq [n] \times [p]$ be a relation between inputs and input \emph{features} of some sort (e.g., key words in text inputs), i.e., $(k,\ell) \in S$ if and only if input $k$ has feature $\ell$. For each $X \subseteq [m]$ let $R|_X$ be the restriction of $R$ to $X \times [n]$, and write $inc(X)$ for the set of inconsistent inputs according to this restriction. For each $X$ and feature $\ell$, let $RS(X,\ell) := \bigwedge S(inc(X),\ell)$, where we indicate logical AND and treat $S$ as a Boolean matrix. Then $RS(X,\ell) = 1$ if and only if every inconsistent file relative to $X$ contains feature $\ell$--in other words, $RS$ measures if feature $\ell$ is responsible for differential behavior. 
	\footnote{
		We might also consider $RS'(X,\ell) := \bigwedge S(inc(X),\ell) \land \bigwedge \lnot S(inc(X)^c,\ell)$, where the complement in $X$ is indicated: this construction additionally requires that every input that is not inconsistent does not contain feature $\ell$.
	}
	
	Writing $\binom{[m]}{r} := \{Y \in 2^{[m]} : |Y| = r\}$, consider $\binom{RS}{\lnot r} := \{\ell : RS\left(\binom{[m]}{m-r},\ell\right) = 1\}$. The sets $\binom{RS}{\lnot r}$ form a partition of $[p]$, and by truncating the range of $r$, we get a principled way to trade off between feature and program coverage. For example, if we consider parsers and white-space delimited tokens, then inputs containing significant natural language (e.g., PDFs with text in streams) will obey Zipf/power law statistics for the frequencies of various tokens, meaning that there is no principled way to truncate the distribution for analysis (e.g., to reverse engineer a particular aspect of the format). The present construction offers a principled workaround that allows us to trade off how many syntax features we consider (along with natural language text) with how many parsers are involved. Moreover, we can build on this by varying $S$ (i.e., taking subsets of features) and comparing the resulting partitions into sets of the form $\binom{RS}{\lnot r}$. By greedily removing features that minimize a distance between partitions (e.g., the variation of information \cite{meilua2007comparing}), we can obtain features that are correlated with program behavior.
	
	\subsection{Related techniques \label{sec:RelatedTechniques}}

	\subsubsection{Bernoulli mixture models \label{sec:BMM}}
	
	\emph{Bernoulli mixture models} (BMMs) provide a natural framework for modeling binary data \cite{saeed2013machine,najafi2020reliable} that has several parallels with TDT. The basic idea of a BMM is to treat a binary matrix as sampled from a probability distribution of the form $\sum_j \alpha_j p_j$ where $\alpha_j \ge 0$, $\sum_j \alpha_j = 1$, and each $p_j$ is a Bernoulli distribution of the form $p_j(x) = \prod_k \pi_{jk}^{x_k} (1-\pi_{jk})^{1-x_k}$. In practice, the number of mixture components (i.e., the range of $j$ above) and the parameters $\pi_{jk}$ are respectively estimated via some model selection criterion and an expectation maximization algorithm. 
	
	With a BMM in hand, we could, e.g., remove small mixture components and renormalize to ``denoise'' the model, but both the model selection criterion and thresholding required here introduce a degree of arbitrariness beyond the modeling \emph{Ansatz}, which may itself be unjustified. Meanwhile, the forensic link between ``noise'' and input behavior is also weaker in the BMM approach than for TDT. That said, it may be useful to consider BMMs in concert with TDT by modeling the initial data as well as the result of removing inconsistent inputs. Such an approach could yield improvements over the approach of \cite{kuchta2018correctness}, which used $K$-means++ clustering of 0-1 vectors encoding errors triggered by files by a given implementation (specifically, of PDF readers).

	\subsubsection{Formal concept analysis \label{sec:FCA}}
	
	TDT is similar in some respects to \emph{formal concept analysis}, not least in that the latter can be viewed algorithmically as the result of a topological simplification of binary data \cite{freund2015lattice,ayzenberg2019topology}. It would be interesting to see if there is some more concrete and direct relationship between TDT and formal concept analysis.

	\begin{acks}
		We thank Sergey Bratus for suggesting that topological properties of set covers ought to inform differential testing. This material is based upon work supported by the Defense Advanced Research Projects Agency (DARPA) SafeDocs program under contract HR001119C0072. Any opinions, findings and conclusions or recommendations expressed in this material are those of the author(s) and do not necessarily reflect the views of DARPA. 
	\end{acks}
	
	\bibliographystyle{ACM-Reference-Format}
	\bibliography{TopologicalDifferentialTesting}
	
	%
	%
	%
	%
	%
	%
	%
	%
	
\end{document}